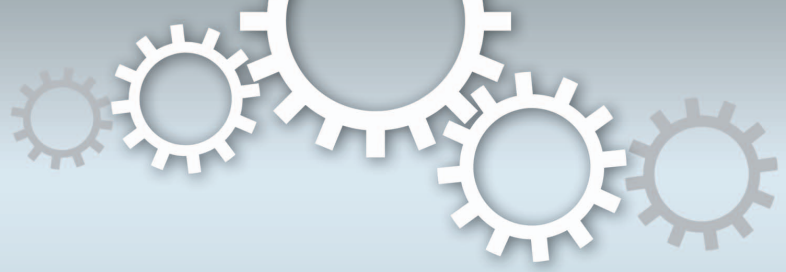

OPEN

# Cooperative protofilament switching emerges from inter-motor interference in multiple-motor transport

David Ando[1], Michelle K. Mattson[2], Jing Xu[1] & Ajay Gopinathan[1]

[1]Department of Physics, University of California, Merced, CA, USA, [2]Developmental and Cell Biology, University of California, Irvine, CA, USA.

Within living cells, the transport of cargo is accomplished by groups of molecular motors. Such collective transport could utilize mechanisms which emerge from inter-motor interactions in ways that are yet to be fully understood. Here we combined experimental measurements of two-kinesin transport with a theoretical framework to investigate the functional ramifications of inter-motor interactions on individual motor function and collective cargo transport. In contrast to kinesin's low sidestepping frequency when present as a single motor, with exactly two kinesins per cargo, we observed substantial motion perpendicular to the microtubule. Our model captures a surface-associated mode of kinesin, which is only accessible via inter-motor interference in groups, in which kinesin diffuses along the microtubule surface and rapidly "hops" between protofilaments without dissociating from the microtubule. Critically, each kinesin transitions dynamically between the active stepping mode and this weak surface-associated mode enhancing local exploration of the microtubule surface, possibly enabling cellular cargos to overcome macromolecular crowding and to navigate obstacles along microtubule tracks without sacrificing overall travel distance.

Conventional kinesin (kinesin-1) is a microtubule-based motor that drives fast and long-range transport of cellular material toward the cell periphery[1]. On the single-molecule level, kinesin is a highly processive motor that can take approximately 100 steps along a bare microtubule before disengaging. Each kinesin has two identical microtubule-binding motor domains ("heads"), which the motor uses alternately to hydrolyze ATP and to step along the microtubule. Mechanisms behind the stepping and processive motion of individual kinesin motors have been studied extensively, with general agreement regarding a head over head mechanism for motors acting by themselves[2,3]. Each kinesin motor has a low sidestepping frequency and typically tracks a single microtubule protofilament during the course of its travel[4]. Perhaps consequently, single kinesin-based transport is highly sensitive to macromolecular crowding on the microtubule surface[5–9].

Intracellular kinesin-based transport is typically accomplished by groups of motors[10,11] that must overcome a highly crowded cellular environment and successfully navigate roadblocks along their microtubule tracks without prematurely dissociating[5]. Defects in kinesin-based transport have been implicated in numerous diseases, especially neurodegenerative diseases[12,13] and quantitative understanding of kinesin's group function is currently an area of active research[12–14]. Clearly, group behaviour can be governed by interactions between motors that are not related to single-motor functions, and these inter-motor interactions must be addressed in experiments employing more than one kinesin per cargo. Recent theoretical and experimental investigations have uncovered evidence for inter-motor interference, and demonstrated that two or more kinesins routinely function via the action of one motor[15,16]. The functional nature of such inter-motor interference is not clear, and has been thus far interpreted as negative interference: when more than one motor is engaged in transport, each kinesin experiences an increased probability of detaching from the microtubule. Intuitively, this effect is negative for group function, since premature detachment of an individual kinesin substantially reduces the travel distance of the group.

Typical efforts to understand function in groups of kinesin motors[16–20] focus on characterizing experimental measurements of the velocity and travel distance of multiple kinesin-based transport[19,21]. However, inter-motor interactions could lead to collective behaviour that manifests itself in other transport characteristics, such as motion perpendicular to the microtubule axis, which requires a more explicit modeling of kinesin properties. A recent study[21] has directly demonstrated such inter-motor interaction, revealing that individual kinesin motors experience an increased likelihood to disengage in active transport while functioning in groups. Experimentally measurements of on axis and off-axis motion of cargo are regularly performed[4,21–24], yet our analysis and experimental observations are unique in their focus on how inter-motor interactions can perturb off-axis motion as motor number and ATP concentration are varied.

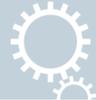

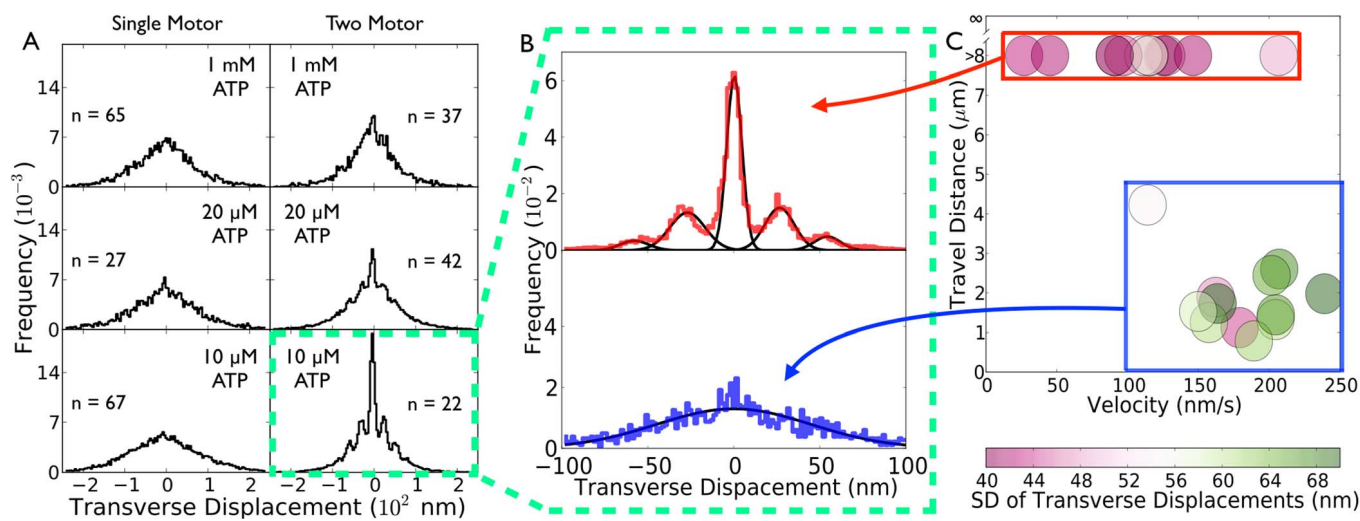

**Figure 1 | Single- and Two-kinesin motor transverse displacement measurements.** (a) Histograms of the transverse displacement of cargos transported by one (left) or two (right) kinesins at the indicated ATP concentrations. DTDs occur at 10 and 20 μM ATP for two-kinesin transport. (b) Top: Histogram of transverse cargo displacement at 10 μM ATP for two motor traces with travel distance >5 μm. Using a binning histogram density of 30 bins per 40 nm, the leftmost sidestep peak is located at -53.3 nm and the remaining sidestep peak on the left side is located at -29.8 nm. The rightmost sidestep peak is located at 53.3 nm and the remaining right single sidestep is located at 26.6 nm. We find that "left" and "right" DTDs occur with roughly equal probability (49% vs 51%). Bottom: Same as (Top) except depicting traces with travel distances <5 μm. The sum of these two histograms is equal to the histogram of DTDs of two-kinesin transport at 10 μM ATP. A sum of five gaussians approximates the two-motor (>5 μm travel) transverse motion histogram well (Top), while the histogram of DTDs for the shorter traces (Bottom) was well approximated by a single gaussian. (c) Two-motor traces at 10 μM ATP separate into two groups. One group (boxed in red) has a low standard deviation in its transverse displacement size of 46 nm on average, has a high processivity (>5 μm), and a low velocity of 109 nm/s on average. Another group (boxed in blue) has traces with a travel distance of >5 μm, a high standard deviation in its transverse displacement size of 62 nm on average, and a higher velocity of 187 nm/s on average.

In this paper, we address collective motor behavior in a controlled manner by utilizing polystyrene beads as an *in vitro* cargo and employing a single antibody to recruit exactly two kinesins onto each bead. The resulting close proximity between kinesins on an identical microtubule mimicked motor arrangements observed for cargos in vivo[25,26]. In contrast to single kinesin's low sidestepping frequency[4], our positional tracking data showed that cargo can be significantly and frequently displaced transverse to the microtubule axis in a *discrete* fashion with no significant loss in processivity. To understand our measurements of group motor transport we used an explicit state-transition model with inter-motor interactions which enabled us to extract the full spectrum of dynamics of individual motors in a group setting, rather than just their average behaviour. Modeling the discrete transverse displacements required the introduction of a surface-associated mode of kinesin in which the motor is not actively stepping, but remains in contact with the microtubule due to the active engagement of other motors. We propose that individual motors in a group setting can utilize a radically different form of stepping across the microtubule surface. Increasing the frequency of kinesin detachment in multiple motor configurations via inter-motor interference and a surface associated state may benefit group function *in vivo* by enabling a group of kinesins to avoid roadblocks along the microtubule, while the enhanced stochastic dissociation and rebinding of individual kinesins in a group can increase the available microtubule landscape surrounding the motors[23].

## Results

**Emergence of discrete transverse displacements (DTDs) in two-kinesin transport.** To probe the functional interactions between two kinesin motors transporting the same cargo, we focused on the changes in cargo position in the direction transverse to the microtubule between subsequent recording frames (33.3 ms temporal resolution). We observed considerable changes in the transverse displacement of two-motor cargos with reducing ATP concentration, but not for single-motor cargos (Fig. 1A, Supplementary Discussion Fig. 1). Specifically, we observed symmetric transverse displacement for both one- and two-motor cargos (Fig. 1, and representative traces in Supplementary Discussion Fig. 1), indicating that cargo motion is not biased in either direction perpendicular to the microtubule. However, with reducing ATP concentration (10 and 20 μM), distinct peaks emerged within the cumulative distributions of the two-motor transverse displacement due to what we term discrete transverse displacements (DTDs). The width and variance of these DTD peaks were substantially smaller than those of the one-motor distribution (>10-fold, Fig. 1A, Supplementary Discussion Fig. 1). The narrow DTD peak width suggests that these peaks correspond to a geometry in which the cargo is linked to the microtubule by two motors, thus increasing the effective linkage stiffness and reducing the effect of thermal noise on cargo position.

**DTDs arise from simultaneous association of both kinesins with the microtubule.** At the lowest ATP concentration tested (10 μM), we observed two distinct classes of transverse displacement in two-kinesin transport at 10 μM ATP, one exhibiting pronounced DTD features, and one exhibiting a broad distribution similar to that observed in single-kinesin transport (Fig. 1B). These two classes of transverse displacement behaviour are well correlated with cargo travel and velocity along microtubules (Fig. 1C), and we observed DTDs only when the cargo also exhibited substantial travel distance and somewhat slower velocity along microtubules (Fig. 1B,C). Observations at higher ATP concentrations followed a similar pattern (Supplementary Discussion 1), so further analysis of DTDs focused on the least noisy 10 μM ATP data.

The observed link between cargo travel distance and the form of transverse displacement is perhaps not surprising; when the cargo is linked to the microtubule via only one kinesin, it experiences greater thermal fluctuation (Fig. 1A, left) and an increased probability of dissociating from the microtubule. Thus, we expected that the single-kinesin bound state dominates cargo trajectories in the group which undergoes limited travel along the microtubule (<5 µM traces). Similarly, as the dissociation rate for each bound kinesin motor is increased at higher ATP concentrations, we expected an increased probability of occurrence of the one-kinesin bound state and an increased presence of the broad, single-motor-like transverse displacement behaviour. These hypotheses were confirmed for two-kinesin transport at higher ATP concentrations (20 µM ATP is shown in Supplementary Fig. 9). On the other hand, at 10 µM ATP, why is the single-kinesin bound state predominant for particular traces during two-kinesin transport? Perhaps in certain cases it is not possible for two motors to simultaneously engage the microtubule in the absence of load due to variations in motor-bead attachment geometry among specific beads. Regardless of the specific mechanism, at 10 µM ATP these single kinesin-like cargo trajectories did not obscure the traces which displayed emergent DTDs and did not significantly impact our quantitative modeling, as described later in the text.

The histograms of the observed DTDs were well approximated by the sum of five Gaussians (Fig. 1B, top, black lines) with the following three important features. First, the width of the central DTD peak (zero transverse displacement) was substantially smaller than that of a single-motor distribution (>10-fold reduction), indicating that the cargo is linked to the microtubule via the stiffer linkage of two kinesin motors. Second, four DTD peaks occurred in two-kinesin transport that were absent from measurements taken for stationary cargos affixed to glass surfaces (Supplementary Fig. 7). This control experiment using beads non-specifically attached to the coverslip demonstrates that DTD events arise from the presence of motors (specifically two motors), and are not artifacts of tracking analysis. Additionally, since these DTD events are not observed in single-kinesin transport, regardless of ATP level used, they are unlikely to arise from experimental error. Nonetheless, by using beads non-specifically attached to the coverslip, we directly demonstrate that DTD events are motor-dependent and must reflect underlying dynamics of the motors interaction with microtubules while bound to the same cargo.

Thus, these DTD peaks are not due to measurement uncertainty (4.5 nm as defined by stationary beads, Supplementary Fig. 7), but arise from the stochastic dynamics of the two available kinesins transporting the same cargo. Additionally, these DTD peaks were uniformly spaced around the central DTD peak, with 27 nm between adjacent DTD peaks. This discrete spacing is 6-fold greater than our measurement uncertainty and is consistent with the hypothesis that each of the two available kinesins stochastically associates/dissociates with/from different protofilaments on the same microtubule during transport, resulting in discrete changes in their common cargo position (see Supplementary Discussion 1 and Supplementary Fig. 8). Taken together, our observations indicate that DTD peaks arise from segments of cargo travel in which the cargo is linked to the microtubule by two kinesins simultaneously. Note that this conclusion is consistent with the observed reduction in the associated cargo velocity (Fig. 1C), which likely arises from negative interactions between the two kinesins which are simultaneously in contact with the microtubule.

**DTD frequency reveals the presence of unbound, surface-associated (SA) motors.** When present, the observed DTDs occurred quite frequently relative to kinesin's individual sidestepping rate. The area enclosed under the four gaussians in Fig. 1B (top, black lines) which have a non-zero mean relative to the total transverse fitted histogram area is representative of the frequency of DTDs, a ratio which indicates that DTDs have a 71.5% probability of occurrence in each two motor trace's frame, or equivalently occur at a rate of at least 21.5 DTD/s (see Methods). First we rule out the possibility that DTD events arise from transitions between two AE (actively engaged) states and one AE state due to inter-motor strain. As demonstrated previously[17], in two-kinesin transport, the transition rate from two-motor bound to one-motor bound states occurs at a rate $\epsilon_2 = 2k_{off} \exp(F/F_d)$, with $k_{off}$ the single motor off rate, $F$ the inter-motor strain, and $F_d$ the detachment force estimated to be ~ 3 pN[17]. If DTDs occur due to transitions between two AE (actively engaged) states and one AE state $\epsilon_2$ is bounded by the DTD frequency of 21.5/s, resulting in an inter-motor strain $F$ of at least 12 pN. This magnitude of inter-motor strain is not likely as it is more than double the single motor force production of kinesin. Thus it is not possible that DTD events arise from unbound motors rebinding during two-kinesin transport.

We further conclude that because DTD-containing traces have a high DTD rate that a mechanism other than AE motor sideways stepping is responsible for producing DTD events. Given that the DTD containing traces have an average velocity of 108.6 nm/s at 10 µM ATP (corresponding to a forward-stepping rate of 13.6 steps/s), if only one motor (either motor at anytime) is actively engaged at all times with an unbound partner motor the engaged motor would have to sidestep on average 158 sidesteps per 100 forward steps (21.5 DTD/s/13.6 steps/s) to generate the observed DTD rate of our cargo. If on the other hand both motors are always continuously engaged, generating the observed average DTD rate from only AE motor sideways stepping would require that each kinesin makes an average of 79 sidesteps per 100 forward steps (21.5 DTD/s/ (2 * 13.6 steps/s)), although some individual traces would show a higher side stepping rate than forward stepping rate even in this extreme case. Given that a single conventional kinesin motor switches protofilaments with a low probability[4] of approximately 13 sidesteps per 100 steps (or 1.6 sidesteps/s at 10 µM ATP), both assumptions require a significantly higher AE motor sideways stepping rate for a single kinesin (>6x) than has been measured experimentally. Due to the high processivity of the DTD-containing traces it is unlikely that the AE sideways stepping rate for individual motors in our two motor transport is significantly higher than has been measured for single motors, we conclude that a mechanism other than AE motor sideways stepping underlies the high DTD rate.

To understand possible mechanisms which could be responsible for this rate we consider a mechanism in which cargo-bound motors can either interact non-specifically with the microtubule or actively engage and drive forward motion, with the particular state of an individual motor being strongly influenced by the states of the partner motor(s) driving the same cargo. This approach departs from the current, more limited treatment in which negative interactions between motors only lead to complete dissociation of the motors from the microtubule and to premature termination of cargo transport. Instead, we consider motors which can be held near the microtubule surface by their partner in a surface associated (SA) state in which the motors weakly contact the microtubule and experience positional constraints on their location due to the action of partner motors on the same cargo. We defined the SA state of kinesin as a kinesin motor whose "head" domains maintain contact (via non-specific interactions) with the microtubule binding lattice while the motor remains unbound and is not performing an ATP-driven process; an AE motor was defined as a motor with at least one of its two heads in an AE state and performing an ATP-driven process. In this SA state, motors diffuse along the microtubule binding lattice, resulting in the rapid and discrete displacement perpendicular to the microtubule of cargos transported by multiple motors. The existence of a SA state has been suggested previously[27–30] to act as a tether influencing the on/off rates of an individual motor dimer. In our study, since we find that DTDs geometrically require only the shifting of a single motor head, as opposed to the entire functional dimer, to an adjacent protofilament (Supplement Discussion 1), we expli-citly allow individual motor heads within each dimer to make contact across different protofilaments in the weakly associated SA state.

 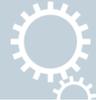

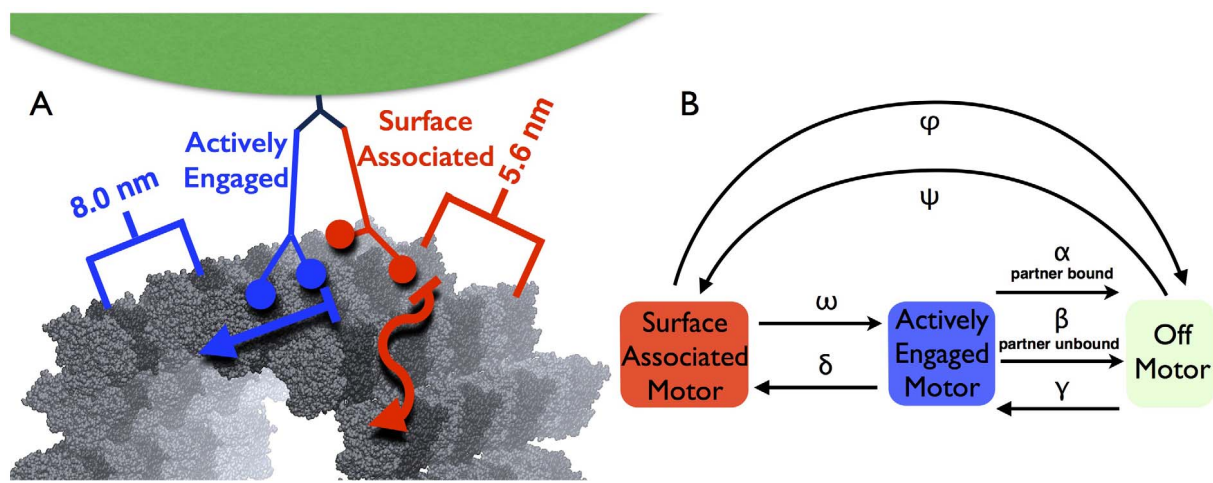

**Figure 2 | Experimental and Model Arrangements.** (a) Our experimental setup consists of a 440-nm diameter polystyrene bead as the cargo (green), with two recombinant K560 half-height kinesins (blue, AE kinesin; red, SA kinesin) connected through C-terminal histidine tags to a single monoclonal antibody (dark blue). Below the motors is shown a 13-protofilament right-handed A-lattice microtubule[39]. The microtubule kinesin binding domain lattice is depicted; dark grey tubulins actively bind kinesin heads, while light grey tubulins interact through non-specific interactions. The transverse distance of 5.6 nm between microtubule protofilaments appears in red, and the axial spacing of 8 nm between kinesin binding locations appears in blue. (b) Transition rate variables and states in our dynamic two-motor state transition model.

Assuming a rigid kinesin-bead complex, where motors are perpendicular to the microtubule surface, the change in measured transverse position of a bead's center of mass can be related to the change in the center of mass of the kinesin motor heads along the microtubule surface by a geometrical projection (Supplement Discussion 1). The size of DTD events, which are measured relative to the change in center of mass of the transported *bead*, is constant within a given trace at around 27 nm in the transverse direction and different traces consistently show this same magnitude for DTD events. Using a geometrical projection we calculate that DTD events with a 27 nm transverse displacement correspond to a 1.4 nm shift in the center of mass of the kinesin motor *heads*: The movement of a single kinesin head (among 4 present in our double motor construct) to an adjacent microtubule protofilament 5.6 nm away would therefore result in the center of mass of all motor heads changing by 1.4 nm, i.e. 5.6 nm/4 = 1.4 nm. We can therefore conclude that the motion of a single motor head moving to an adjacent protofilament imposes a magnitude of transverse motion on the transported bead equal in size to those of DTD events, indicating that DTD events likely originate from protofilament switching of individual motor heads.

Our measurements using smaller cargos (200 nm diameter) further support the validity of the geometrical interpretation in our model and the resulting potential for a SA state being responsible for the high DTD rate. If the geometry of the motor-cargo association is correctly described by Eq. 1 of Supplementary Discussion 1, then the transverse fluctuations for this smaller cargo will reduce to approximately 59% of that observed in our main experiments with 440-nm beads. Indeed, under otherwise identical conditions (20 $\mu$M ATP), we observed substantially more transverse fluctuations in two-kinesin transport for the larger bead. The standard deviation in the observed transverse fluctuation decreased from 41.7 nm to 23.4 nm when we reduced the bead diameter from 440 nm to 200 nm, in excellent agreement with the value of 24.7 nm predicted by our model (Supplementary Fig. 6). Thus, unbound, SA motors seem able to contribute to the transport of intracellular cargos.

**Stochastic state transition model with an SA state reproduces the observed transverse motion and travel distance.** The experimental data for processivity and transverse displacement rate (Fig. 1, Xu et al[19]) are replicated by a generic kinesin transition rate model that includes the SA state and interference between motors. We explicitly simulate the two-motor system, dynamically modelling interference and the state of each motor throughout time; each motor can be in the AE state, the SA state, or the off state (completely unbound from the microtubule) (Fig. 2). The motors were considered to be permanently attached to the cargo bead but could change their state of attachment to the microtubule over time with seven transition rates and the state of the partner motor (Fig. 2B). We recorded the transverse and axial positions of the motors during the simulation with the axial position set by active ATP stepping of either motor and transverse displacements determined by the AE sidestepping rate and SA hopping rate $\epsilon$ (see Methods).

The undetermined transition rates in our model were fitted by best matching simulated processivity values at all ATP concentrations to experimentally measured processivity in two-kinesin transport[19] and by simultaneously matching simulated transverse displacement histograms to experimental histograms at 10 $\mu$M ATP (see Methods). Using a brute-force search over a wide parameter space followed by a conjugate gradient search, a set of well performing rates which minimized the $\chi^2$ error of the model to experiment were determined (Table 1). Optimal parameter values resulted in a model which closely matched our experimental results (Fig. 3, Table 1) with a $\chi^2$ fit of 0.2.

Using the optimal fitted transitions rates (Table 1), our model demonstrates that during two-motor travel, only a relatively small fraction of travel consists of both motors being in the AE state simultaneously: 26.3% of travel at 10 $\mu$M ATP, 25.6% of travel at 20 $\mu$M ATP, and 13.5% of travel at 1 $m$M ATP. On the other hand, the SA state is able to capture a significant fraction of multi-motor behavior, depending on the ATP concentration; 46.4% of travel at 10 $\mu$M ATP, 32.4% of travel at 20 $\mu$M ATP, and 14.3% of travel at 1 $m$M ATP has one motor in a SA state and the partner motor in an AE state. Although a relatively small fraction of two-motor steps may consist of simultaneously engaged AE motors, these steps may take longer on average than single-motor steps due to inter-motor strain between AE motors, resulting in DTD-containing traces at 10 $\mu$M ATP being slower than the non-DTD traces, as we observed. SA motors may also provide some resistance in the axial direction of movement, which may also contribute to the reduced velocity of these DTD containing traces. Indeed, other experiments revealed

Table 1 | Overview of model parameters, optimal model values, and experimental measurements of kinesin-based transport. Parameters are defined in Figure 2b

| Parameter | State transition rate | Optimal model values [1 mM ATP], (Experimental value) |
|---|---|---|
| $\beta$ | AE → OFF rate (if partner not AE) | 0.32 $s^{-1}$, (1 $s^{-1}$)[40,41] |
| $\gamma$ | OFF → AE rate | 6.4 $s^{-1}$ (5.0 $s^{-1}$)[42] |
| $\alpha$ | AE → OFF rate (if partner AE) | 23.1 $s^{-1}$ |
| $\omega$ | SA → AE rate | 45.6 $s^{-1}$ |
| $\delta$ | AE → SA rate | 17.3 $s^{-1}$ |
| $\phi$ | SA → OFF rate | 2.2 $s^{-1}$ |
| $\psi$ | OFF → SA rate | 2.2 $s^{-1}$ |
| $\epsilon$ | SA protofilament switching rate | 63.7 $s^{-1}$ |

reductions in velocity as motor numbers in kinesin-based transport increased[31,32].

## Discussion

Our experiments were carried out *in vitro*, in the absence of external load[19]; such that at any instant in time, the cargo is linked to the microtubule via either a one- or two-kinesin bound state[17]. When thermal motion is factored out, the transverse displacement of the cargo is dominated by the stochastic binding/unbinding of each available kinesin to the microtubule, since each kinesin typically tracks a single microtubule protofilament during transport[4,24]. In the presence of thermal motion (our experiments were conducted at room temperature), the cargo's transverse displacement also reflects the stiffness of the cargo's linkage to the microtubule through the motors. Since two kinesins should stiffen this linkage, we predicted that thermal motion would contribute less to the cargo's transverse displacement when two motors were linked to the cargo than when a single motor provides the linkage. Since each kinesin associates longer with the microtubule at lower ATP concentrations (e.g. 10 and 20 $\mu$M)[19], we interpret our experimental observation of DTD peaks as resulting from a reduced ATP concentration which increases the duration of cargo binding in the two-kinesin or one-kinesin bound state. This yielded slow enough dynamics for us to cleanly probe the stochastic dynamics underlying cargo transport by two kinesins. Note that for single-kinesin transport, the cargo can only be linked to the microtubule via a single motor at any instance, and thus we observe cargo transverse displacement that remains unperturbed by varying the ATP concentration given that thermal motion dominates observed transverse cargo displacement in this case.

As sideways stepping by AE motors alone is unlikely to be responsible for the high DTD rate we observed, this rate is most likely due to interference between motors and a mechanism in which motor heads are associated with the microtubule surface via non-specific interactions and diffuse along the surface. However other explanations are possible such as a mechanism in which the motor-cargo recruitment geometry undergoes relatively rapid switching between conformations. We feel that it is improbable that DTDs are due to conformational changes because DTDs result in preferred absolute positions for the centre of mass of the cargo (Supplementary Fig. 8). While up to five peaks are visible in the histogram of per-frame cargo displacements at 10 $\mu$M ATP for two-kinesin transport, some of these traces exhibit up to nine distinct peaks in the histogram of absolute transverse position of the cargo (Supplementary Fig. 8). Peaks in the absolute transverse position histogram were spaced by approximately 27 nm, close to the average transverse differential sidestep size measured. It is unlikely that multiple conformational changes in the kinesins-linker complex are of approximately the same magnitude, or that changes in motor conformation results in effects on cargo position that are identical to the effects of motors moving between protofilaments. Rather, peaks in the absolute transverse position histogram likely correspond to different attachment geo-metries of the two kinesin motors to the microtubule surface (Supplementary Fig. 5). Furthermore, given the geometry of two-motor transport, each peak in the absolute transverse position histogram corresponds to the projection of a single motor head moving to an adjacent protofilament (Supplementary Discussion 1), and DTDs which are 27 nm in size correspond to single motor heads moving to adjacent protofilaments. This geometrical interpretation also explains why only up to 9 peaks can be seen in the histogram of absolute transverse position (Supplementary Fig. 4,5).

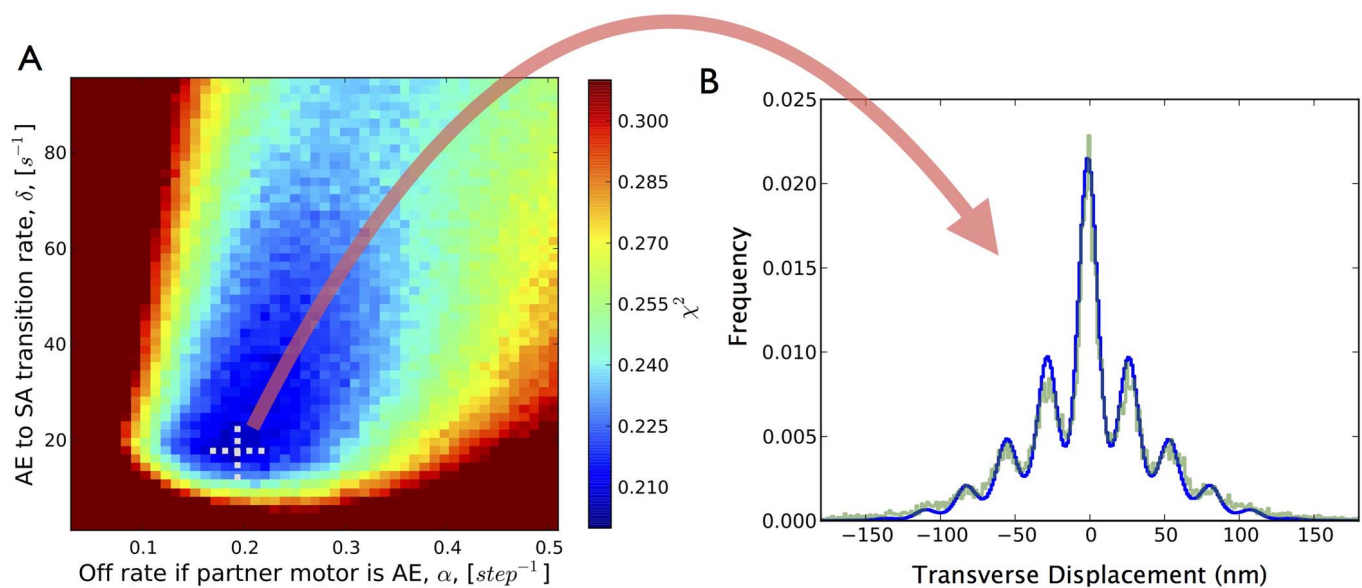

**Figure 3 | Model Results.** (a) $\chi^2$ error over the undetermined off rate and the AE-to-SA state transition rate phase space. Our model best fits experimental data at an AE to SA transition rate of 17.3/s and an off rate of 0.195/step when the partner motor is in the AE state. There appears to be a well behaved single global $\chi^2$ error minimum. (b) Comparison of the histogram of transverse differential stepping for experimental traces at 10 $\mu$M ATP versus modelled transverse motion at the best-fit model parameters.

Alternatively, various types of interference between motors may underlie the frequent generation of DTDs. Inter-motor strain or steric interactions between motors may result in greatly enhanced sidestepping rates for individual motors. However this type of interactions would increase the motor's off rate and significantly decrease processivity, which would be incompatible with our observations of large processivities for traces with frequent DTDs. For other types of interference which result in motors rapidly binding and unbinding from the microtubule, we would expect to frequently observe sidesteps of the whole motors themselves, instead of the individual motor heads, as described in Supplementary Discussion 1. This scenario is ruled out by the recruitment geometry in our experiments and by our observation that most DTDs corresponded to single motor heads moving to an adjacent protofilament. Thus a form of interference which reduces the likelihood that two partner motors can both be actively engaged at the same time, but allows one motor to be 'dragged' along the microtubule is sufficient to model observed experimental results. Such interference could physically result from the attachment geometry of the motors to the cargo or from the motor's conformation. This type of negative interference between AE kinesins was previously demonstrated *in vitro* by Diehl[16], and is incorporated into our theoretical model.

We assumed that motor heads in the SA state weakly interact with the microtubule binding lattice via non specific interactions, which implies that AE and SA motors moving in the transverse direction are observationally indistinguishable in our experiment. This is because both situations can give rise to the same 27-nm cargo displacements via a geometrical projection driven by moving one kinesin motor head in the transverse direction by 5.6 nm, the transverse distance between microtubule protofilaments. The discordance between previous experimental measurements of AE sidestepping and the high frequency of sustained sidesteps we observed in the two-motor case suggests that the vast majority of DTDs are due to SA motor heads changing position from one protofilament to an adjacent one 5.6 nm away in the transverse direction. Direct measurement of the transverse displacement of an individual motor head within a kinesin dimer (5.6 nm) would be challenging. Here, single-head movement was resolved by projecting the position of kinesin heads onto the position of the transported cargo.

Our experiments have yielded the first resolved measurements of the interactions of multiple motors with individual microtubule protofilaments, revealing that the simplistic view of individual kinesins as either in the AE state or completely off the microtubule is insufficient to explain our observations. This is likely due to having two kinesin motors which are bound to the cargo in nearly identical locations while sharing the same microtubule, which should maximize interference between motors. Consideration of the off-axis component of cargo motion uncovered a new SA state for motors which indicates that inter-motor interference during multiple motor-based transport does not simply lead to premature dissociation of motors, but may increase motor-cargo-track association by leading to a 'tether'-like SA state, and that one motor can drive the majority of motion in multiple motor transport while maintaining high processivity. Thus, transverse displacement can be a powerful experimental tool for directly probing the underlying dynamics of kinesins driving the same cargo.

Enhanced transverse motion may contribute to the flexibility needed for multiple motors to navigate cargos through crowded cellular environments (including obstacles along the microtubule surface) without losses in travel distance. Increased transverse motion in multiple motor complexes may also increase the rate at which proteins bind cargos which are in the process of being transported and facilitate the binding of kinesins which are bound to the same cargo yet remain unattached to a microtubule. Further, the tether-like SA state may be functionally similar to the gliding behaviour of the dynactin tail over the microtubule surface, which functionally increases the travel distance of dynein[33].

Previous investigations have suggested that only one motor is active for the majority of the time in multiple motor transport[15,34]. Derr *et al*[34] reported a linear increase in travel distance as the kinesin copy number increased, instead of the exponential increase expected in models in which motors are either on or off in the conventional mean field interpretation. Furuta *et al*[15] determined that force production for multiple motors plateaued at two-motor force production irrespective of the number of engaged motors. Both of these investigations indicated that multiple motors interfere with each other and may frequently force each other into the SA state. The SA state may also be important in reducing the number of motors in the AE state in multiple-motor transport to minimize ATP hydrolysis, as an energy efficiency or conservation mechanism.

The specific geometry of the attachment of the motors to the cargo bead enabled our observation of a high effective sidestepping rate for the cargo. Different geometries could result in very different behaviours, such as a small range of motion for the cargo (shorter distance from motor to cargo) and possibly no SA sidestepping. Thus, the location of the kinesin-binding domains on the cargo could constitute an important regulatory mechanism for transport, allowing the cell to fine-tune the type of motion and/or processivity required for transporting a given cargo to its target.

Given that conventional kinesin has functionally evolved to transport cargo in a multiple motor setting, it may be inappropriate to consider kinesin stepping mechanisms devoid of the influence of other motors, with new mechanisms and motor states possibly representing a more fundamental way in which to understand kinesin function, rather than through the properties of single motor-cargoes which are rarely encountered *in vivo*.

## Methods

**In vitro motility experiments.** Protein purification and details of motility experiments are as described previously[19]. Briefly, for two-kinesin motility experiments, histidine-tagged recombinant kinesin K560[35] was specifically recruited to each polystyrene bead via penta-His-antibody (0.44 $\mu$m diameter, Qiagen[36]). Casein (5.55 mg/mL) was utilized to block antibody-independent, non-specific binding of motors onto beads. The incubation ratio of penta-His antibody to polystyrene bead was titrated down to the single-antibody-per-bead range to ensure that the maximum number of recruited kinesins per bead was two. Optical trap-mediated force measurements were employed to specifically select the population of beads carried by two active kinesins. For multiple-kinesin experiments where the number of kinesins was more than one but not well defined, kinesin was directly adsorbed onto carboxylated beads (200 nm diameter, Polysciences). The single-molecule range for both motor-cargo recruitment methods was determined via Poisson statistics[22,37]. For the stationary bead control experiment, bare beads (0.44 $\mu$m diameter, not coated with kinesin) were adsorbed non-specifically onto a coverslip in 35 mM PIPES. The optical trap was shut off for all motility measurements to ensure measurements of bead travel without external load. Bead motility along the microtubule was imaged via differential interference microscopy[37], with video recorded at 30 fps (Basler), digitized using Lagarith lossless compression for subsequent data analysis.

**Data analysis.** Video recordings of bead motion were particle-tracked to within 10-nm resolution (1/3 pixel, see also Supplemental Fig. 7) using a previously described template matching algorithm[22,38]. To decouple bead motion parallel to, and transverse to the microtubule, individual trajectories of bead motion were projected along the microtubule lattice via least square regression analysis.

DTD determination was performed by first histogramming the transverse differential motion between all frames in traces which were over 8 $\mu$m long. This resulting histogram was well matched by the sum of five gaussian curves, Fig. 1B, with non-zero centered gaussians representing DTD events as typically the histogram of transverse differentials consists of a single zero centered gaussian curve. The 21.5 DTD/s constraint for two-motor transport at 10 $\mu M$ ATP was calculated by first performing a multi-gaussian fit to the histogram of transverse differentials, calculating the ratio of the area under the four fitted gaussians which had non-zero mean relative to the total histogram area, and multiplying this ratio by the experimentally used observation frame rate to constrain the minimal rate of DTD events per second (21.5 DTD/s = 0.715 * 30/s frame rate).

**Model.** We explicitly modelled the two-motor system by allowing each motor to be in the AE state, the SA state, or the off state during any given step and by recording the transverse and axial positions of the motors during the simulation. For simplification, absolute transverse motion was allowed to remain unbounded in the simulation (only differential transverse motion was compared between model and experiment). The stepping rate for the motors (modelled motor velocity) was fixed by the experimentally measured velocity at each modelled ATP concentration[19]. The motors were allowed to change their state of attachment to the microtubule once each step of the motors via transitions rates $\alpha$, $\beta$, $\gamma$, $\delta$, $\omega$, $\psi$, and $\phi$ (Fig. 2B), while changes in transverse position were determined by rate $\epsilon$ and a previously measured AE sidestepping rate[4]. Specifically, we modelled the effect of the geometrical attachment constraints of the motors to cargo which result in intermotor interference by splitting the off rate from the AE state to the off state into two different rates (Fig. 2B). Consistent with previous modelling efforts, these two off rates are assumed to occur due to a per-step physical process, which implies that they will be sensitive to the ATP concentration. We additionally assume that the partner of a SA motor is not allowed to transition to the SA state or to the off state because a SA motor does not hydrolyze ATP and thus cannot actively exert force or strain gate other motors during transport. Transitions from the AE state to the SA state, from the SA state to the off state, and from the off state to the SA state occur as a per-second diffusive search-like processes and are independent of the ATP concentration. Processivity histograms at all 3 ATP concentrations were assumed to have the same percentage of DTD to non-DTD containing traces, a 44/56 ratio, as found at the 10 $\mu M$ ATP concentration, with non-DTD traces modelled as having only one motor that contacts the microtubule. Cargo bead trajectory simulations were terminated when both motors were in the off state.

The histogram of modelled transverse displacement at 10 $\mu M$ ATP and the processivity of model trajectories at 1 $mM$, 10 $\mu M$, and 20 $\mu M$ ATP were compared to their experimentally measured equivalents; optimal $\alpha$, $\beta$, $\gamma$, $\delta$, $\omega$, $\psi$, $\phi$, and $\epsilon$ rates which minimized the $\chi^2$ error of the model to experiment were determined by a conjugate gradient search (Table 1).

### Acknowledgments

This work was partially supported by National Science Foundation (NSF) grants EF-1038697 (to A.G.) and NSF-DBI-0960480 (to A.G.), by a James S. McDonnell Foundation Award (to A.G.), and by American Heart Association grant 825278F (to J.X.). The authors would like to thank Steve Gross, K.C. Huang and David Quint for valuable input.

### Author contributions

D.A., J.X. and A.G. conceived and designed the project. J.X. and M.K.M. performed experiments. D.A. and J.X. analyzed data. D.A. performed the simulations. D.A., J.X. and A.G. wrote the main manuscript text and D.A. prepared figures. All authors reviewed the manuscript.

### Additional information

**Supplementary information** accompanies this paper at http://www.nature.com/scientificreports

**Competing financial interests:** The authors declare no competing financial interests.

**How to cite this article:** Ando, D., Mattson, M.K., Xu, J. & Gopinathan, A. Cooperative